\documentclass[usegraphicx]{mn2e}
\def\H0{{\it H}$_0$}

\def\q0{{\it q}$_0$}
\def\kmps{km~s$^{-1}$}

\def\ergps{erg~s$^{-1}$}

\def\nH{$N_{\rm H}$\thinspace} 

\def\psqcm{cm$^{-2}$}
\def\ergpspsqcm{erg~cm$^{-2}$~s$^{-1}$}

\def\cps{ct\thinspace s$^{-1}$}

\def\phpspsqcm{ph\thinspace s$^{-1}$\thinspace cm$^{-2}$}

\title[X-ray nebula around NGC4388] 
{The X-ray nebula around the Seyfert 2 galaxy NGC4388} 
\author[K. Iwasawa et al] 
{\parbox[]{6.5in} {K. Iwasawa$^1$, A.S. Wilson$^2$, A.C. Fabian$^1$ and A.J. Young$^2$}\\
\\
$^1$Institute of Astronomy, Madingley Road, Cambridge CB3 0HA\\ 
$^2$Department of Astronomy, University of Maryland, College Park, MD 20742-2421, USA\\
}
\date{}

\begin{document}

\maketitle

\begin{abstract}
  We report on X-ray emission from the Seyfert 2 galaxy NGC4388
  observed with the Chandra X-ray Observatory. A hard X-ray peak is
  found at the position of the active nucleus suggested by optical and
  radio observations. Extended soft X-ray emission correlates well
  with the ionization cone found in optical line emission. A large
  soft X-ray extension is found up to 16 kpc (and possibly 30 kpc) to
  the north of the galaxy. Photoionized gas with low ionization
  parameters (log $\xi\leq$ 0.4) appears to be the likely explanation
  of this emission. The same ionized gas clouds could be responsible for the
  optical [OIII] emission. Fe K$\alpha $ line emission from cold
  material is found to be extended by a few kpc.
\end{abstract}


\section{introduction}

NGC4388 ($z=0.00842$) is a nearly edge-on ($i\simeq 78^{\circ}$)
spiral galaxy hosting a Seyfert 2 nucleus (Phillips \& Malin 1982;
Filippenko \& Sargent 1985), located near the core of the Virgo
cluster. The recession velocity measured for this galaxy (2540 \kmps)
is much higher than the cluster mean velocity (1100 \kmps). This
suggests that the galaxy is moving away from us through the
intracluster medium (ICM) at a supersonic speed, and an interaction
between the galaxy and ICM has been suspected, particularly in the
context of ram-pressure stripping of the interstellar medium
(Chamaraux et al 1980; Giovanelli \& Haynes 1983; Kenney \& Young
1986; Petitjean \& Durret 1993; Veilleux et al 1999).

Radio images show a compact double central source and a plume
extending to the north (Stone, Wilson \& Ward 1988; Hummel \& Saikia
1991; Falcke, Wilson \& Simpson 1998). No broad emission component was
found in optical spectropolarimetric observations of the
nucleus (Kay 1994). However, the
somewhat controversial detection of off-nuclear broad H$\alpha $
emission by Shields \& Filippenko (1988) led to the idea of an
obscured Seyfert 1 nucleus in NGC4388. Unambiguous confirmation of the
hypothesis came from hard X-ray observations by which a strongly
absorbed (\nH $\sim 10^{23}$\psqcm) X-ray source has been detected
(Hanson et al 1990; Iwasawa et al 1997; Forster, Leighly \& Kay 1999;
Bassani et al 1999).

The presence of extended optical emission-line nebulae in NGC4388 has
been known for a few decades (Ford et al 1971; Sandage 1978; Colina et
al 1987; Pogge 1988; Corbin, Baldwin \& Wilson 1988; Veilleux et al
1999; Yoshida et al 2001). Besides the low excitation optical emission
extending along the galaxy disk, which probably traces active star
forming regions, the high excitation extended emission above the
galactic plane has been of great interest in connection with its
kinematics and ionization mechanism.  Until recently, the extraplanar
ionized gas was known to extend up to 4 kpc from the nucleus,
but the wide-field images obtained from the SUBARU SuprimeCam show
optical filaments extending up to 35 kpc to the north-east (Yoshida et
al 2001).  Studies of the excitation conditions in this extraplanar
nebula suggest that it is likely to be photoionized by the central
active nucleus, although some contribution by shock excitation has
been claimed recently (Ciroi et al 2003). The required ionizing
luminosity is estimated to be of the order of $10^{43}$\ergps (Colina
1992; Kinney et al 1991; Yoshida et al 2001).

The soft X-ray emission of NGC4388 was first shown to be extended from
a ROSAT HRI observation (Matt et al 1994). However, the details of the
extended X-ray emission and its origin have been unclear. In this
paper, we present an imaging observation with the Chandra X-ray
Observatory (Weisskopf et al 2000), which reveals the detailed morphology
of the extended X-ray nebula and spectral variations across the
nebula, as well as the active nucleus seen in hard X-rays. The
distance of the galaxy is assumed to be 16.7 Mpc throughout this
paper. The angular scale is then $\sim 81$ pc arcsec$^{-1}$ (or 12
arcsec corresponding to 1 kpc).

\section{Observation and data reduction}

NGC4388 was observed with the Chandra ACIS detectors on 2001 June 8.
The nucleus of the galaxy was positioned on the ACIS-S3
detector. The focal plane temperature for this observation was
$-120^{\circ}$C. The data reduction was carried out using the CIAO 2.2
package and calibration files in CALDB version 2.10. 

The data at the position of the nucleus of NGC4388 are affected by
photon pile-up (the estimated pile-up fraction is about 10 per cent).
The detector background of ACIS-S3 during this observation was
relatively stable. The analysis for the emission at small radii was
performed using the data of the full exposure of 20 ks. However, for
extended emission found at larger radii, we filtered the event file
using the 2.5--7 keV light curve from a source-free region on the S3
chip. Data taken in periods when the background counts deviate by more
than 20 per cent from the mean value have been discarded, which
results in an exposure time of 14 ks for analysis. The spectral
analysis was performed using XSPEC version 11. The time dependent
degradation of the ACIS sensitivity at low energies have been
corrected with {\tt acisabs} by Chartas \& Getman.

The total background-corrected observed fluxes are estimated to be
$3.4\times 10^{-13}$\ergpspsqcm\ in the 0.5--2 keV band, and
$3.7\times 10^{-12}$\ergpspsqcm\ in the 2--7 keV band.

\section{Active nucleus}


\begin{figure}
\centerline{\includegraphics[width=0.38\textwidth,angle=0,keepaspectratio='true']{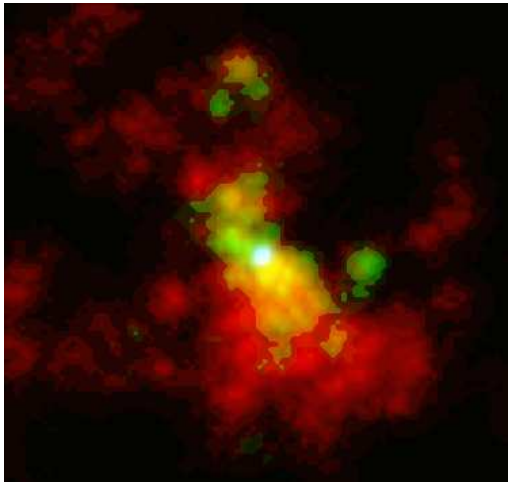}}
\caption{
  A composite X-ray image of the central part ($45\times 43$
  arcsec$^2$, or $3.6\times 3.5$ kpc$^2$) of NGC4388. The image has
  been constructed from three energy bands (Red: 0.3--1 keV; Green:
  1--3 keV; Blue: 4--7 keV). The active nucleus is located at the
  centre of the image, where the hard X-ray emission is peaked. }
\end{figure}

The composite X-ray image of the central part of NGC4388, produced
from the three energy bands, 0.3--1 keV,


1--3 keV and 4--7 keV, is
shown in Fig. 1. The active nucleus is located at the position of the
hard X-ray peak indicated in blue. The soft X-ray emission shows
complex morphology. Particularly bright is a conical feature to the
south. These soft X-ray features are discussed in detail in the
following section. 

\subsection{The hard X-ray nucleus}


\begin{figure}
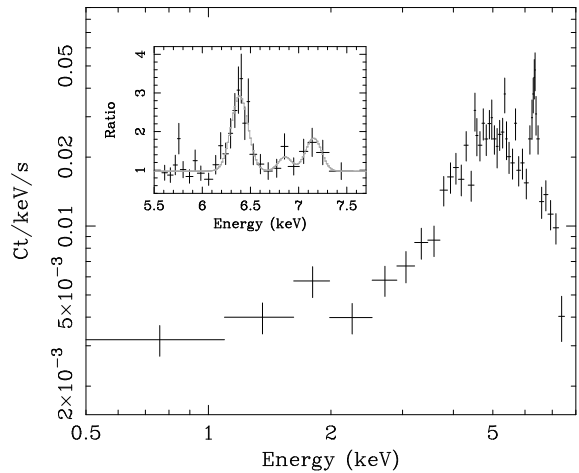

\centerline{\includegraphics[width=0.35\textwidth,angle=270,keepaspectratio='true']{fig2main.ps}}
\vspace{-5.7cm}
\leftline{\hspace{1.7cm}\includegraphics[width=0.14\textwidth,angle=270,keepaspectratio='true']{fig2inset.ps}}
\vspace{3.5cm}
\caption{Main panel: the Chandra ACIS-S spectrum of the nucleus of NGC4388,
  extracted from a region within 1.5 arcsec of the nucleus. The data
  are distorted by mild pile-up. A strong hard excess with an Fe K line
  is seen above 4 keV. The absorbing column density is estimated to be
  \nH $\simeq 3.4 \times 10^{23}$\psqcm, when $\Gamma = 1.8$ is
  assumed. Note that there is a faint, flat spectrum component below 4
  keV. Inset: details of the Fe K band data in the form of a ratio of
  data and absorbed power-law model. The energy scale has been
  corrected for the galaxy redshift. The strong cold Fe K$\alpha $
  at 6.4 keV and weaker higher-energy line features are seen. The
  grey-line shows the best-fit model including three gaussians,
  converted to the ratio form.}
\end{figure}

The active nucleus of NGC4388 is very faint below 4 keV in the Chandra
image and not visible below 1 keV, due to obscuration probably
associated with gas in the edge-on stellar disk.

Strong emission is seen at energies above 4 keV. Its spectral
shape is consistent with a strongly absorbed source (Fig. 2). The
spectrum presented in Fig. 2 was taken from within 1.5 arcsec radius
of the hard X-ray peak. It can be broadly described as having a
heavily absorbed power-law above 3 keV with Fe K line features at 6.4
and near 7 keV and a flat component at low energies. As mentioned in
Section 2, the nucleus is so bright that the data were affected by
moderate photon pile-up (estimated to be $\approx 10$ per cent). The
main consequences of the photon pile-up are a hardening of the
spectrum and a reduction in the detected flux. We correct for these
effects using the pile-up model of Davis (2001).

The absorption column density is found to be \nH $\simeq
3.5^{+0.4}_{-0.3}\times 10^{23}$\psqcm (hereafter quoted errors are
for a 90 per cent confidence range for one parameter of interest,
unless stated otherwise), when the photon index of the power-law
continuum is assumed to be 1.8 and a simple absorbed power-law is
fitted to the 4--7.8 keV data. A deep Fe K photoelectric absorption
edge seen above 7.1 keV is consistent with the large absorption column
density.

A strong Fe K$\alpha $ line is found at an energy of
$6.36^{+0.02}_{-0.02}$ keV (hereafter quoted line energies are
corrected for the galaxy redshift) with an equivalent width of
$EW=440\pm 90$ eV. The line flux is $(9.3\pm 1.9)\times
10^{-5}$\phpspsqcm. This line is unresolved and the 90 per cent upper
limit of the line width is FWHM $\leq 23500$ \kmps. There is also an
excess peak at $7.1\pm 0.1$ keV with $EW = 200\pm 120$ eV, and a less
significant peak at $\simeq 6.8$ keV with $EW\simeq 60$ eV (90 per
cent confidence detection). The estimated line fluxes for these lines
are $\simeq 2.7\times 10^{-5}$ and $\simeq 1.2\times 10^{-5}$\phpspsqcm,
respectively. Given the low pile-up fraction, these weak emission
features (at least the statistically robust one at 7 keV) are unlikely
to be artifacts of pile-up. A similar feature was detected at $6.9\pm
0.1$ keV in the ASCA SIS spectrum. Fe K$\beta $ emission is predicted
to be three times weaker than the observed value. It could be blended
with FeXXVI K$\alpha$ at 6.97 keV from a highly ionized medium.  A
similar K$\alpha $ line from hydrogen-like iron has been detected in
the XMM-Newton spectrum of the Seyfert 2 galaxy Mrk463 (Sanders et al
2003 in prep.). The presence of such a highly ionized Fe line is
consistent with the high ionization gas (log $\xi\sim 3$) hypothesized
in the following section.

Extrapolating the absorbed power-law leaves a faint excess below 4
keV. The spectrum is nearly flat $\Gamma = -0.3^{+0.5}_{-0.6}$ in the
0.5--2 keV band. This component is probably scattered light of the
hidden active nucleus. As the soft X-ray image shows, the nucleus is
very faint below 1 keV, suggesting further obscuration to the
scattering region, in addition to the primary obscuration of the
central source occuring probably at much smaller radii.

The observed fluxes (corrected for pile-up) in the 0.5--2 keV and 2--7
keV bands are $2.1\times 10^{-14}$\ergpspsqcm\ and $2.7\times
10^{-12}$\ergpspsqcm, respectively. The estimated absorption-corrected
2--10 keV luminosity, excluding the iron line luminosity, is
$6.3\times 10^{41}$\ergps. This
value is lower than the two previous ASCA observations in 1993 and
1995 (Iwasawa et al 1997; Forster et al 1999) by a factor of 2--3.

\subsection{Position of the active nucleus}

The obscuration of the nuclear region affects the measurements of the
position of the optical nucleus of this galaxy. The northern blob of
the central double radio source has been considered to be the active
nucleus, since it has a flat spectrum (Carral, Turner \& Ho 1990). The
best registration of the radio and optical images obtained so far is
the one by Falcke et al (1998, see the paper for detailed discussion)
for the VLA and HST data. The positions of the HST red peak and the
northern blob in the VLA 3.5 cm map are in agreement to within 0.45
arcsec. Our 4--7 keV image shows a strong point-like source with a
faint envelope. In this energy band, the nucleus is largely free of
obscuration and its position should coincide with the active nucleus.
We have applied the latest aspect correction using the alignment files
released in 2002 May by the Chandra Science Center.  With this aspect
correction, the expected absolute positional accuracy is smaller than
0.6 arcsec.  Our 4--7 keV peak is located at R.A. =
$12^{\rm h}25^{\rm m}46^{\rm s}.77$, Dec. =
$+12^{\circ}39^{\prime}44^{\prime\prime}.0$ (J2000), which is $\approx
0.6$ arcsec NE of the VLA radio nucleus. Given the uncertainties, the
optical, radio and the X-ray positions are all in agreement.

\section{Soft X-ray nebula}


\begin{figure*}
\hbox{\hspace{8mm}{\includegraphics[width=0.45\textwidth,angle=0,keepaspectratio='true']{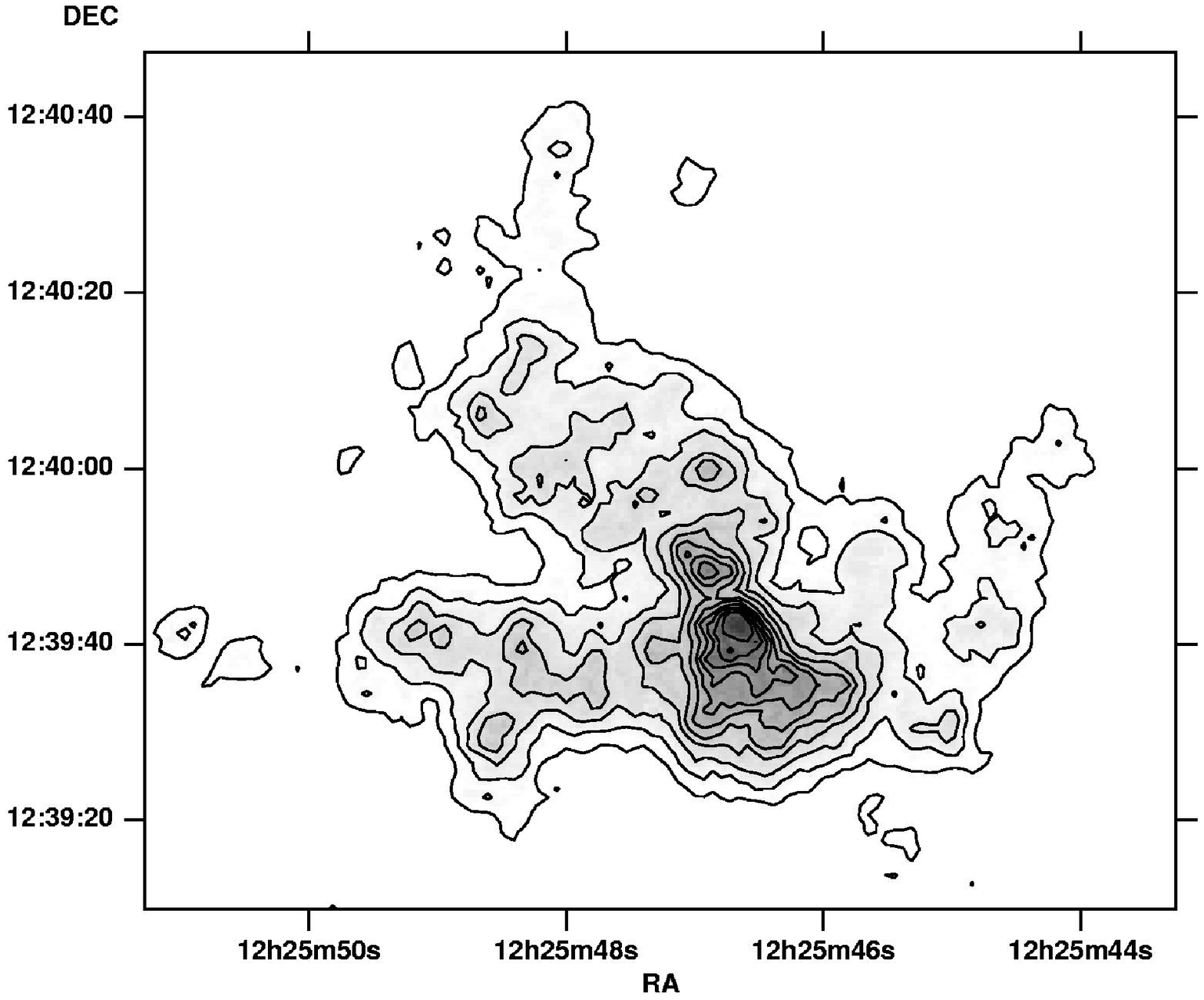}}\hspace{7mm}{\includegraphics[width=0.45\textwidth,angle=0,keepaspectratio='true']{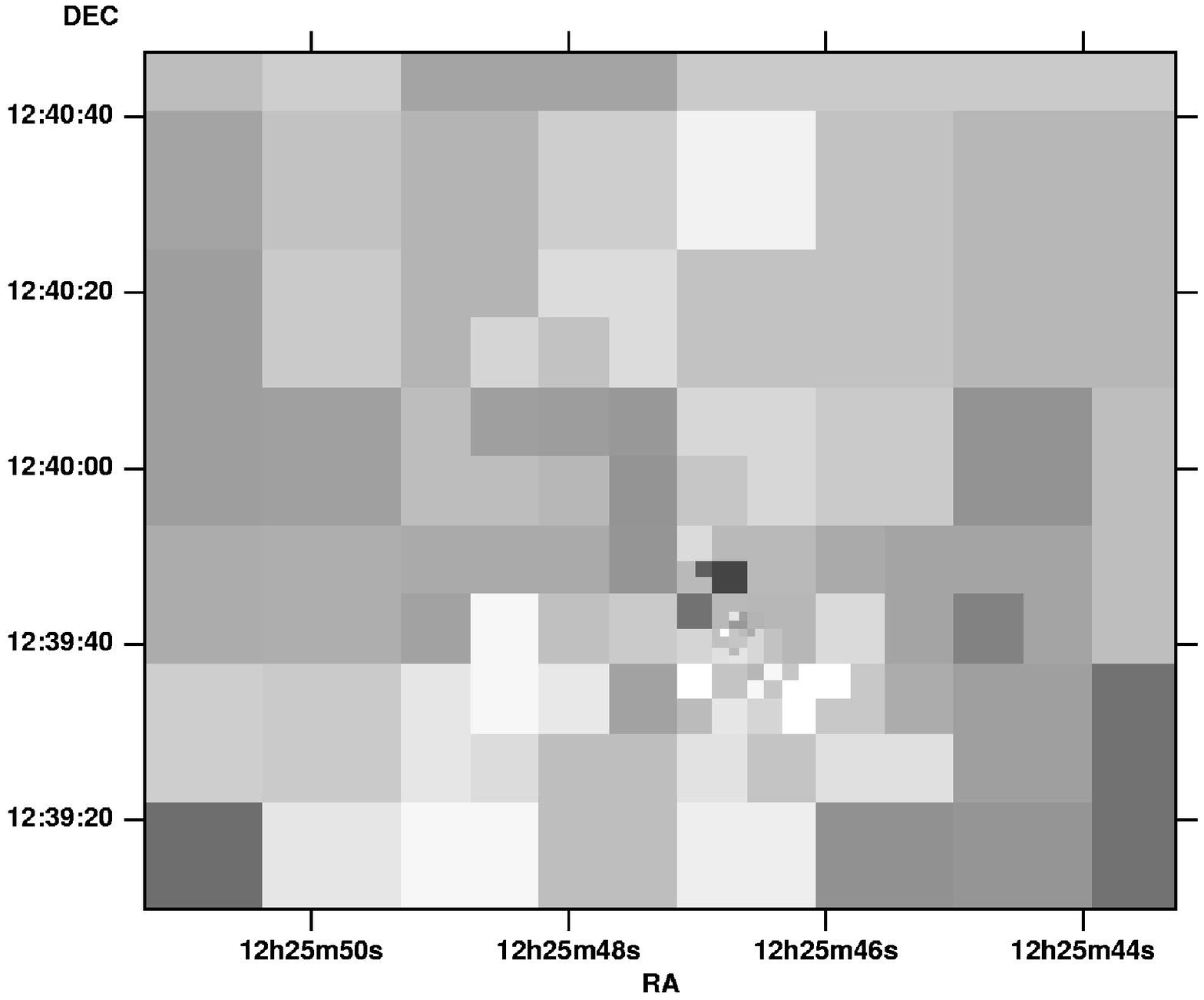}}}
\caption{
  a) Left panel: The soft X-ray (0.3--1 keV) image of NGC4388
  ($2^{\prime}\times 1.65^{\prime}$: 9.6 kpc $\times$ 8.0 kpc). The
  sky coordinates are of J2000. The contours are drawn at ten
  logarithmic intervals in the range of 0.5--50 per cent of the peak
  flux, which occurs at the inner part of the southern cone. b) Right
  panel: The hardness ratio image of the same field as the left panel.
  Count rate ratios, C(0.7--1 keV)/C(0.4--0.7 keV), are indicated
  on a logarithmic scale between 0.01 and 10 (the darker the shade,
  the larger the count rate ratio). The hardest spectrum occurs at the
  north plume near the nucleus. The three-shell region to the
  north-east (Section 4.3) has a relatively hard spectrum. The
  southern cone and the eastern extension along the stellar disk show
  a softer spectrum on average while there is evidence for some
  variations within the region.}
\end{figure*}

The soft X-ray emission is spatially extended and shows interesting
morphology. The 0.3--1 keV image of the central $2^{\prime}\times
1.65^{\prime}$, or $9.7\times 8.0$ kpc is shown, along with the X-ray
colour map (0.7--1 keV / 0.4--0.7 keV) in Fig. 3. Some regions of
interest are investigated separately below. Basic data for these
regions are summarized in Table 1.

\subsection{Southern cone}

The soft X-ray (0.3--1 keV) image shows a well-defined conical
extension to the south (Fig. 1, Fig. 3a), resembling the [OIII]$\lambda 5007$
ionization cone (Pogge 1988; Falcke et al 1998; Veilleux et al 1999;
Yoshida et al 2001). The western side of the soft X-ray cone, in
particular, has a sharp, straight boundary at $PA\approx 220^{\circ}$.
Its apex is, however, slightly north of the hard X-ray
peak. With the apex of the cone slightly displaced from the nuclear
position (0.5 arcsec to the east, 2.5 arcsec to the north) when
extrapolated with the rims of the cone, the opening angle of the X-ray
cone is measured to be $\approx 55^{\circ}$.

The surface brightness of the soft X-ray emission drops at the centre
due to strong absorption, whilst at radii from the nucleus larger than
2 arcsec, where the brightness peaks, its profile shows an exponential
decline ($\propto \exp[-(r-2.6)/3.9]$, where $r$ is the radial distance
from the hard X-ray nucleus in arcsec).  An alternative description is
with a King profile with a core radius of 5.7 arcsec (or 460 pc) and
a slope index of $-1.7$.  In either case, the soft
X-ray emission reaches the background level at radii of 20--25 arcsec
(1.6--2 kpc), and there appears to be no further emission at larger
radii to the south (in contrast to the northerly direction; see the
following section).

The ACIS-S spectrum of the southern cone is shown in Fig. 4. The
spectral data were taken from a conical region in the radial range of
1.5--19 arcsec from the nucleus. The spectrum is dominated by soft X-ray
emission below 3 keV.

Although the steep rise of the soft X-ray spectrum down to 0.5 keV can
be approximated by a power-law form with photon index of $\Gamma\sim
3.4$, the most likely explanation for the soft X-ray emission is a
blend of strong emission lines. Modelling by thermal emission spectra
(from collisionally ionized plasma, e.g., MEKAL, Kaastra 1992) is not
favoured on the following grounds. If the observed X-ray emission is
due to collisionally ionized plasma, a likely origin is a galactic
outflow driven by a nuclear starburst. Circumstantial evidence against
this hypothesis is that, although star formation is taking
place in the spiral arms of the stellar disk (Pogge 1988), there is no
strong evidence for a compact nuclear starburst in NGC4388. To drive a
well-collimated outflow, as seen in the southern cone, a compact
starburst region needs to be thermalized to form a high pressure core
(e.g., Chevalier \& Clegg 1985), which does not appear to be the case
in NGC4388.

Spectral analysis does not support the thermal emission
hypothesis either. A single temperature model fails to explain the strong
OVII emission lines at 0.56 keV, even with non-solar abundance ratio.
Strong OVII is emitted from $kT\simeq 0.15$ keV gas if it is
collisionally ionized. Since this low temperature gas does not emit
much above 0.7 keV, a higher temperature gas is required to
explain the data. A two-temperature model in which one component has a
fixed temperature of $kT_1=0.15 $ keV and the other has a fitted
temperature of $kT_2 = 0.53^{+0.07}_{-0.07}$ keV gives a reasonable
fit to the data up to 1.5 keV with $\chi^2 = 30.1$ for 39 degrees of
freedom. This fit, however, underestimates the Mg emission feature at
1.3 keV by a factor of 2, and the Si feature at 1.8 keV by a factor of
4, on extrapolation. This modelling also has two significant problems: 1) the
metallicity, which is assumed to be identical between the two
components, is very low, $0.08^{+0.14}_{-0.06}$ solar. This is not
consistent with the gas which should have been enriched by supernovae
through a starburst; 2) large internal absorption of \nH
$=2.4^{+1.6}_{-1.5}\times 10^{21}$\psqcm\ is required. With the
Galactic dust-to-gas ratio, this implies visual extinction of $A_{\rm
V}\simeq 2.2$ mag. Unlike the northern extension (see Section 4.2),
the southern cone is located well outside the obscuration and such a
large extinction is unlikely to be present (even the nuclear optical
spectrum taken from inner radii shows a smaller reddening $A_{\rm
V}\simeq 0.9$, Petitjean \& Durret 1993). 

We therefore explore the possibility of photoionized plasma, as
postulated for the optical ionization cone. In terms of quality of the
fit to the data, the photoionization model described below gives a
better fit to the data in the same energy range (0.4--1.5 keV) with
$\chi^2 = 19.1$ for 40 degrees of freedom.

High resolution X-ray spectra of photoionized gas in Seyfert 2
galaxies obtained from grating spectrometers show many emission
features from a broad range of ionization in the soft X-ray band
(e.g., see Kinkhabwala et al 2002; Brinkman et al 2002 for XMM-Newton
RGS and Chandra LETGS data, respectively, of NGC1068; Sako et al 2000 for
Mrk 3). These emission lines are heavily blended together and mostly
difficult to resolve at the spectral resolution of the CCD. This heavy
line blending also makes an estimate of the underlying continuum
difficult in a CCD spectrum, if it is present at all. Therefore, we
will not perform a detailed analysis here as one would accomplish with
a high resolution grating spectrum, but instead only describe some key
features characterizing the observed spectrum and present a possible
model, which should be treated as a guide only.

A few of the strongest emission-line peaks can be recognized in the
data at 0.56, 0.86, 1.33, 1.77 and 2.33 keV, for which the typical
error in the line centroids is 0.04 keV and the significance of the
detection is larger than 90 per cent confidence level. These
correspond to the OVII triplet, a blend of various Fe L emission,
OVIII Radiative Recombination Continuum (RRC) and NeIX (0.92 keV), and
low ionization (IV--VIII; for the ionization parameter inferred below)
lines of Mg, Si and S, respectively. These are all low ionization
features, and the 0.4--0.9 keV data and the three discrete
emission-lines above 1 keV are explained well with emission from
optically thin gas with log $\xi\simeq 0.4$ (here, we compare with
model spectra generated by XSTAR version 2.1 by T. Kallman, see
Kallman \& Bautista 2001, for optically thin gas, illuminated by a
power-law continuum between 13.6 eV and 50 keV with a photon-index of
1.8. The ionization parameter is defined as $\xi = L/(nR^2)$ erg cm
s$^{-1}$, where $L$ is the 1--1000 Ryd luminosity, $n$ is the density
and $R$ is the distance from the ionizing source. XSTAR calculates
physical conditions for a spherical gas shell by solving radiative
transfer. The quoted value for $\xi$ is for the inner edge of the
shell. Therefore the calculated spectra include emission from a range
of $\xi$ lower than the quoted value from outer radii). This fit
requires no significant excess absorption by cold gas above the
Galactic value (\nH $= 2.6\times 10^{20}$\psqcm, Dickey \& Lockman
1990). The only significant residual below 0.9 keV is an excess at
0.65 keV, which could be OVIII Ly$\alpha $, indicating the presence of
higher ionization gas.

The 0.9--3 keV data unexplained by the above ${\rm log}\thinspace\xi\simeq 0.4$
emission, apart from the Mg, Si and S emission lines, may be emission
and scattered continua from highly ionized, low density gas, as
suggested by the excess OVIII at 0.65 keV. Unfortunately, there is no
other significant detection of spectral signatures, which could be
used to constrain the ionization parameter. If FeXXV at 6.7 keV,
possibly present in the hard X-ray spectrum (see Section 5.1), is real
and originates from the same medium as the OVIII emission, then the
ionization parameter $\xi $ would be a few thousands. For example, the
spectrum from high ionization gas with $\xi = 2000$, when combined
with the low ionization (log $\xi = 0.4$) spectrum, can provide a
reasonable fit to the 0.4--3 keV data (see Fig. 4). This high
ionization spectrum has broader radiative recombination continua due
to high temperature of the gas ($T\sim 10^6$ K), as well as high
ionization lines, which could fill the continuum between the emission
lines of low ionization spectrum. A possible alternative which would
play the same role is the soft X-ray excess component of the
reflection spectrum, which is seen in the hard X-ray band (see e.g.,
Ross, Fabian \& Young 2000 for a computed reflection spectrum from
mildly ionized matter). Further investigation is, however, beyond the
capability of the present CCD spectrum.

In summary, low ionization (log $\xi\simeq 0.4$) photoionized gas
appears to be a plausible explanation for prominent soft X-ray spectral
features; Another component, which could be highly ionized (log
$\xi\simeq 3$) gas, is required, although its origin is not clear. It
should, however, be noted that this two-component photoionization
model is only an approximation required to describe the spectrum at
the CCD resolution with limited signal-to-noise ratio, from which only
a few strong emission features can be recognized. Emission features
from photoionized gas with a broad range of ionization parameter are
likely to be present and would be detected in a high resolution spectrum.

There are significant spectral changes within the southern cone (see
Fig. 3b). Spectra from radial ranges of 1.5--7 arcsec (0.1--0.6 kpc)
and 7--19 arcsec (0.6--1.5 kpc) are compared (cf. the [OIII]/H$\alpha
$ maximum occurs at 3 arcsec from the nucleus, Falcke et al 1998;
Veilleux et al 1999). At the low energy end, the spectrum from the
larger radii is slightly softer than that from the inner part. The
spectral softening at larger radii can be explained by a slight
increase in ionization parameter for the low ionization gas, or a
decrease in absorption, or both. In the second case, with a constant
ionization parameter of log $\xi = 0.4$, a marginal excess absorption
(\nH $\approx 2\times 10^{20}$\psqcm ) above the Galactic value is
found for the inner part while no excess absorption is required for
the outer part.


\begin{figure}
\centerline{\includegraphics[width=0.55\textwidth,angle=270,keepaspectratio='true']{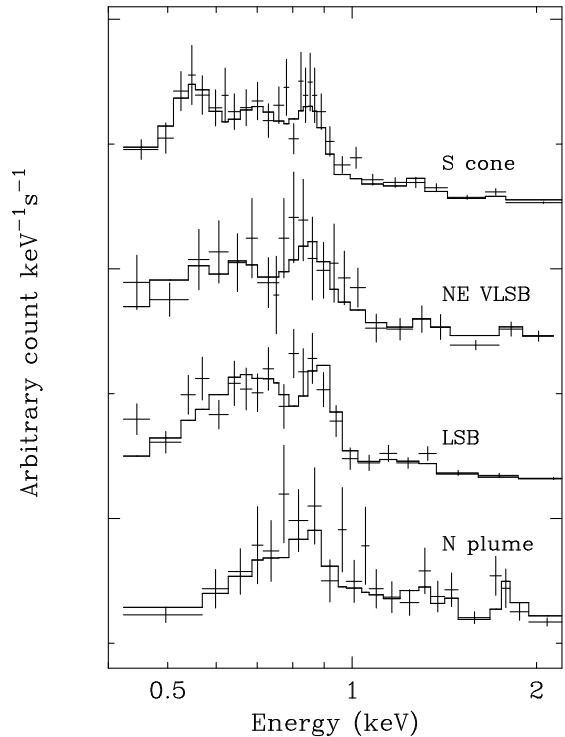}}
\caption{
  The soft X-ray spectra (the y-axis has a linear scale) taken from
  the southern cone, the north-east very low surface-brightness (NE
  VLSB) emission, the low surface brightness (LSB) emission within 4
  kpc and the northern plume within 1 kpc (from top to bottom).
  Information on each region is summarised in Table 1. To compare the
  relative strengths of spectral features in the spectra, they have
  been normalized to the emission peak at 0.8--0.9 keV and an
  arbitrary offset added for clarity. Note the strong OVII line at
  0.55 keV in the S cone spectrum and the absorbed flat spectrum for
  the N plume. The solid-line histograms indicate possible models of
  emission from photoionized gas, for which details are given in the
  text.}
\end{figure}

\subsection{Northern extension within 1 kpc}


\begin{figure}
\centerline{\includegraphics[width=0.42\textwidth,angle=0,keepaspectratio='true']{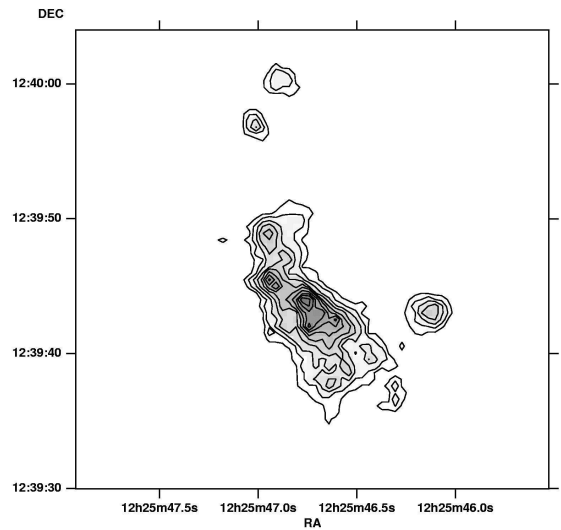}}
\caption{
  The 1--2 keV image of the central part of NGC4388. The active
  nucleus is located at the central peak. The contours denote nine
  logarithmic intervals in the range of 0.15 to 80 per cent of the
  peak brightness. }
\end{figure}

\begin{figure}
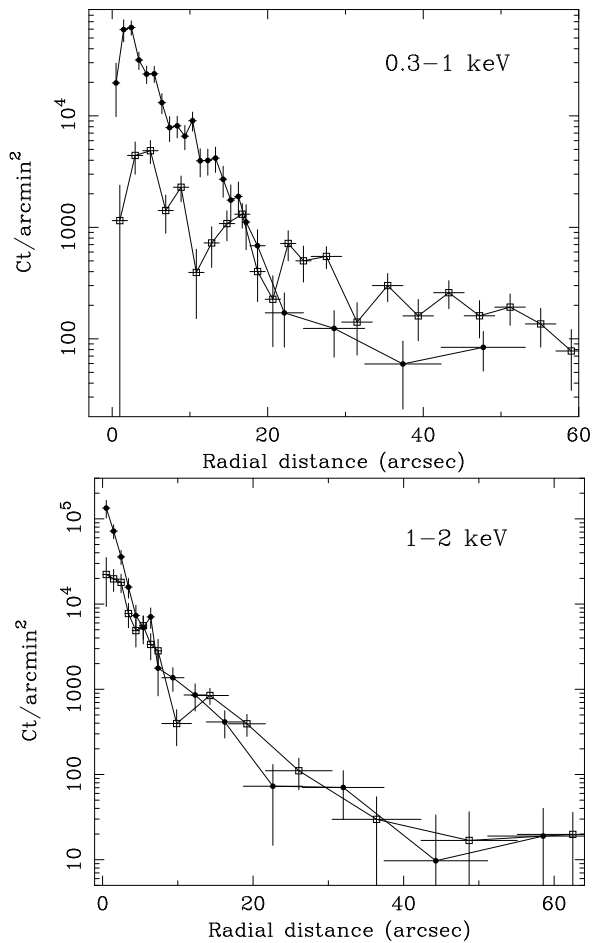

\centerline{\includegraphics[width=0.35\textwidth,angle=270,keepaspectratio='true']{fig6a.ps}}
\centerline{\includegraphics[width=0.35\textwidth,angle=270,keepaspectratio='true']{fig6b.ps}}
\caption{The surface brightness profiles of the NE (Open squares,
PA=355--75$^{\circ}$) and SW (Filled circles, PA=165--225$^{\circ}$)
cones in the 0.3--1 keV (upper panel) and 1--2 keV (lower panel)
bands.}
\end{figure}

In contrast to the southern extension, the inner part on the north
side of the nucleus (within $\sim 15$ arcsec, or
1.2 kpc) is faint in the soft X-ray (0.3--1 keV) band.  However, the
appearance of the image changes dramatically in the higher energies.
In the 1--2 keV band, the northern and southern extensions show
comparable brightness, apart from at very inner radii (Fig. 5). This
is also demonstrated by radial profiles in the two energy bands (Fig.
6).

This change in appearance of the X-ray image can be understood as a
result of obscuration toward the northern extension (and the nucleus).
The soft X-ray spectrum taken from the radial range of 3--11 arcsec
(0.2--0.9 kpc) to the north of the nucleus is shown as ``N plume'' in
Fig. 4. In comparison with the southern cone spectrum, the soft X-ray
excess below 0.9 keV is much less pronounced, and the low ionization
emission lines from Mg and Si in the 1--2 keV range are stronger
relative to the emission complex at lower energies. The data can be
modelled by the same combination of low and high ionization gases as
used for the southern cone spectrum (see Section 4.1) but with a
larger absorption column density of \nH = $2.1^{+0.6}_{-0.5}\times
10^{21}$\psqcm (the model is shown in solid histogram together with
the data in Fig. 4).  Combining with the radial surface brightness
profiles in Fig. 6, it can be argued that, in the absence of
absorption, the nebulae to the north and south would share similar
brightnesses and possibly ionization conditions.

NGC4388 is viewed nearly edge-on, with the near side of the stellar
disk being tilted upward by $\sim 12^{\circ}$ (see Fig. 5 of Veilleux
et al 1999).  This enables us to have a clear view of the SW cone. On
the other hand, the tilted near side of the disk blocks our line of
sight towards the northern extension, suppressing low energy X-ray
emission through photoelectric absorption. Probably we see the
northern extended X-ray emission through the stellar disk for which
the inferred column density appears to be reasonable.

There are a few morphological and spectral features to note on the
northern extension: the 1--2 keV image (Fig. 5) shows that a
relatively narrow bright filament emanates from the nucleus at
$PA\approx 55^{\circ}$ before bending northwards at $\sim 3$ arcsec
from the nucleus, past which point the filament opens up, similar to
the radio plume imaged with the VLA (Falcke et al 1998).

\subsection{Lower surface brightness emission within 4 kpc}

Besides the bright southern cone and the northern extension at small
radii, lower surface brightness soft X-ray emission is seen in the
radial range of 1--4 kpc from the nucleus. The extension is to the
east along the stellar disk, to the west with a bent morphology, and
to the north-east, as shown in Fig. 3a. Although there is possible
evidence for spectral variations across the low surface brightness
emission (Fig. 3b), limited statistics due to the short exposure of
our observation prevent us from dividing the low surface brightness
emission into multiple regions and analysing their spectra separately.
The soft X-ray spectrum integrated over this region is shown in Fig.
4 as ``LSB''.

Although, as cautioned above, the photoionization model should be
taken as only a guide, the spectrum favours a higher ionization
parameter of log $\xi\simeq 1.4$ for the low ionization gas than that
for the southern cone. A large absorption of \nH
$=2.5^{+0.9}_{-0.8}\times 10^{21}$\psqcm\ is required. It should also
be noted that the emission extended along the stellar disk might be
different in origin.  As the optical H$\alpha $ image and the
excitation map (Corbin et al 1988; Veilleux et al 1999; Yoshida et al
2001) show, this region is the site of intense star formation in the
spiral arms, and the soft X-ray emission could originate in thermal
hot gas with a temperature of a few million K.  The spectrum there
lacks the strong OVII feature at 0.55 keV, but the quality of the
present data cannot provide a clear answer to the question of whether
the gas is collisionally- or photo-ionized. 

To the north-east, the soft X-ray image shows three shell-like
structures, aligned parallel to each other with an approximate
separation of 10 arcsec, or 0.8 kpc (Fig. 7). They are located in the
radial range of 14--40 arcsec (1.1--3.2 kpc) from the nucleus at PA$\simeq
40^{\circ}$, and coincide with the location of the [OIII] north-east
complex (e.g., Veilleux et al 1999). The colour map (Fig. 3b) shows that
the three-shell region has a harder spectrum than the surroundings. In
fact, the shell structure is clearer in the 0.7--1 keV image than in
the 0.4--0.7 keV image.


\begin{figure}
\centerline{\includegraphics[width=0.25\textwidth,angle=270,keepaspectratio='true']{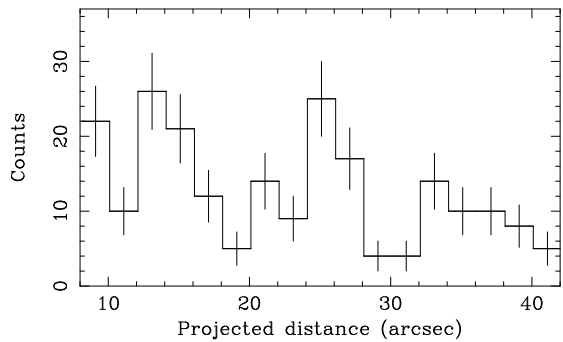}}
\caption{
  The projected surface brightness profile of the three shells seen in
  the 0.3--1 keV image (Fig. 3) to the north-east. The x-axis denotes
  the angular distance from the nucleus in P.A. =
  38$^{\circ}$. The detected counts are integrated over a 25-arcsec
  wide strip for each bin (the bin width is 2 arcsec). Three peaks are
  seen at around 14, 25 and 35 arcsec in projected distance.}
\end{figure}

\subsection{Northern extension beyond 4 kpc}


\begin{figure}

\centerline{\includegraphics[width=0.55\textwidth,angle=0,keepaspectratio='true']{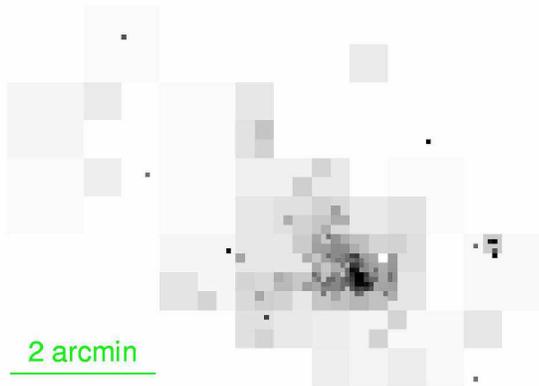}}
\caption{
  The large-scale soft X-ray (0.3--2.5 keV) image of NGC4388. The
  background-subtracted 5-arcsec bin image has been adaptively binned
  to show faint, large scale structures. Dark dots scattered over the
  region are mostly point sources.}
\end{figure}

The large-scale image in the 0.3--2.5 keV band is shown in Fig.
8. The original image was produced from the entire ACIS-S3 chip with 5
arcsec binning. In this energy band, the Virgo cluster emission is
significant. The background
obtained from a source-free region on the same detector was assumed to
be constant over the field. After subtracting the background, the
image has been adaptively binned so that each pixel has a
signal-to-noise ratio larger than 2.5.
 
Very low surface brightness emission extends to the north-east
beyond the region shown in Fig. 3. There is a low sensitivity belt on
the detector due to the CCD node boundary, which extends in
$PA\approx 120^{\circ}$ at around 1.8 arcmin ($\sim 9$ kpc) from the
nucleus. Including this sensitivity gap, extended emission with
significance above $3\sigma $ reaches 3.3 arcmin ($\sim 16$ kpc) from the
nucleus. It coincides with the region where emission-line filaments
are seen in both the H$\alpha $ and [OIII] images taken by the SUBARU
Suprime-Cam (Yoshida et al 2001).

The radial surface brightness profile of the 0.3--2.5 keV emission to
the north-east (in the P.A. range between $0^{\circ}$ and
$50^{\circ}$) can be fit well with a power-law ($\propto r^{\alpha}$;
$r$ in kpc) with $\alpha = -1.39\pm 0.06$. A fainter excess is seen
further out at PA$\sim 45^{\circ}$, 4 arcmin (20 kpc) away from the
nucleus, where there is a bright complex of H$\alpha $ knots, which is
a part of the SUBARU 35-kpc-long filament (Yoshida et al 2001).
However, the significance of this X-ray excess is just above
$2\sigma$.

The spectrum taken from the region between 4 kpc and 9 kpc in radius
is shown in Fig. 4 (as ``NE VLSB''). No significant emission is detected
above 2 keV.  Two prominent
emission lines from low ionization Mg and Si at 1.3 and 1.8 keV,
respectively, are seen in the 1 -- 2 keV range. The inferred ionization 
parameter is found to be slightly low at
log $\xi\simeq 0.3$ with Galactic absorption.

\begin{table*}
\begin{center}
\caption{
  Selected regions for analysis of extended X-ray emission. Energy
  spectra for these regions are presented in Fig. 4 and Fig. 10. S
  cone: a conical region to the south of the nucleus in the radial
  range of 1.5--19 arcsec from the nucleus; N plume: a region of
  1.5--12 arcsec to the north of the nucleus; LSB: low
  surface-brightness region consisting of the NE three-shell region
  (see Section 4.3), the galactic ridge to the east, and the western
  ridge seen in Fig. 3; and VLSB NE: a region to the north-east for
  the very low surface-brightness emission (see Fig. 8) in the radial
  range of 50--110 arcsec. The fluxes given below are as observed
  (corrected only for the efficiency degradation of the ACIS
  detector).}
\begin{tabular}{lcccc}
Region & Area & Count rate & $F_{\rm 0.5-2keV}$ & $F_{\rm 2-4keV}$\\
& arcmin$^2$ & $10^{-2}$\cps & \ergpspsqcm & \ergpspsqcm \\[5pt]
S cone & 0.062 & $4.02\pm 0.14$ 
& $11.4\times 10^{-14}$ & $2.9\times 10^{-14}$\\
N plume ($r\leq 1$ kpc) & 0.021& $1.09\pm 0.07$ 
& $2.5\times 10^{-14}$ & $2.4\times 10^{-14}$\\
LSB ($r\approx $1--4 kpc) & 0.58 & $3.17\pm 0.15$ 
& $8.3\times 10^{-14}$ & $2.0\times 10^{-14}$\\
VLSB NE ($r\approx $4--9 kpc) & 1.10 & $1.32\pm 0.14$ 
& $4.2\times 10^{-14}$ & --- \\
\end{tabular}
\end{center}
\end{table*}

\section{Hard X-ray emission}

\subsection{Circumnuclear hard X-ray emission}


\begin{figure}
\centerline{\includegraphics[width=0.42\textwidth,angle=270,keepaspectratio='true']{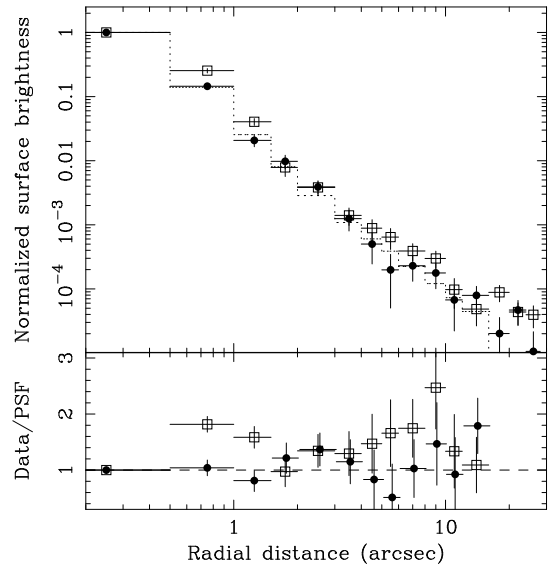}}
\caption{
  Upper panel: surface brightness profiles of 4--7 keV emission taken
  from PA ranges of 255$^{\circ}$--80$^{\circ}$ (open squares) and
  80$^{\circ}$--255$^{\circ}$ (filled circles). The radial distance is
  measured from the position of the hard X-ray nucleus. The innermost
  bins of respective profiles have been corrected for pile-up (by 10
  per cent), and the data are normalized to the corrected innermost
  bins. The dotted-line histogram shows a simulated PSF of
  monochromatic 6.4 keV X-rays (note that the PSF profile is truncated at
  15 arcsec). Bottom panel: ratios of the data of the two radial
  profiles to the PSF. The profile for the southern P.A. region is
  consistent with the PSF while that of the northern P.A. region is
  broader than the PSF.}
\end{figure}


\begin{figure}
\centerline{\includegraphics[width=0.55\textwidth,angle=270,keepaspectratio='true']{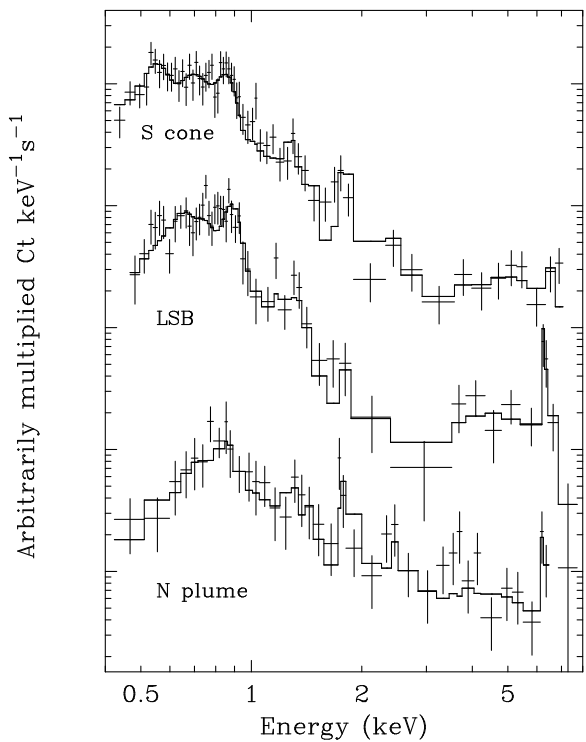}}
\caption{
  Full-band ACIS spectra (both x- and y-axes have logarithmic scales)
  of the southern cone, low surface brightness emission within 4 kpc,
  and the inner north plume (from top to bottom). All the three
  regions shows weak hard X-ray tails in their spectra. It should
  however noted that the hard tail in the S cone spectrum could be
  largely due to the scattered wing of the nuclear source PSF (see
  Section 5). No significant hard X-ray emission above 2 keV is
  detected at radii larger than 4 kpc.  Solid-line histograms show
  fitted models including a reflection continuum (and a Fe K line
  where applicable) in the hard band in addition to those for
  photoionized gas emission for the soft X-ray band, presented in Fig.
  4. Note the strong Fe K line at 6.4 keV in the LSB spectrum in
  contrast to the lack of such a strong line in the southern cone
  spectrum.}
\end{figure}

The 4--7 keV emission is highly concentrated around the nucleus
(approximately 90 per cent of the total 4--7 keV flux comes from the
central 2 arcsec in radius) but marginally resolved. The surface
brightness distribution of the hard X-ray emission is found to be
skewed slightly to the north at low brightness levels (see also the
following subsection on the Fe K$\alpha $ emission). Although the
asymmetry suggests that the extension is real, since the point spread
function (PSF) of the Chandra optics (HRMA) has a broad low-level wing
at high energies, we have examined the radial distribution of the hard
X-ray emission.  Fig. 9 shows radial surface brightness profiles in
the 4--7 keV band, taken from northern and southern regions divided at
P.A. of 80$^{\circ}$ and 255$^{\circ}$ to compare with the PSF (NB.
the northern region overlaps the southern cone slightly at its inner
western edge). We note that the photon pile-up at the nucleus results
in a slightly diluted core. The innermost bins of the two observed
profiles have been increased by 10 per cent as an approximate
correction for pile-up.

A PSF was simulated at the monochromatic energy of 6.4 keV for the
same position on the detector as that of the hard X-ray nucleus, using
the standard PSF library in CIAO. The total hard X-ray spectrum peaks
at the Fe K$\alpha $ band (6.1--6.6 keV), which carries $\simeq 20$
per cent of the observed 4--7 keV counts. Because the off-axis angle
of the source is small (0.65 arcmin) and 6.4 keV is close to the
higher end of the bandpass, the simulated PSF can be regarded as a
conservative estimate of the PSF for the 4--7 keV emission.
While the radial profile for the southern half is consistent with the
PSF, the northern half shows significant excess emission above the PSF
but reaches the background level at around 30 arcsec ($\sim 2.5$ kpc). 

The origin of the extended hard X-ray emission is probably some form
of reflection of the hidden active nucleus, as suggested by the hard
spectrum (Fig. 10). This appears to be true, at least, for the
spectrum of the low surface brightness emission region which shows a
strong Fe K line at $6.35\pm 0.06$ keV with $EW=1.8\pm 0.9$ keV. A
similar picture is consistent with the noisy spectrum of the inner
northern plume, which barely shows evidence for an Fe K line with
$EW=770\pm 610$ eV. This cold reflection could contribute to low
ionization lines at lower energies (e.g., 2--4 keV band), depending on
the ionization state and obscuration to the reflecting medium. The
reflecting medium could be molecular clouds distributed in the stellar
disk (recall the Sgr B2 clouds near our Galactic centre, which is
located 100 pc away from the Sgr A$^{\star}$, e.g., Koyama et al
1996). If ram-pressure stripping of the gas in the host galaxy is
taking place (e.g., Petitjean \& Durret 1993; Cayatte et al 1994;
Veilleux et al 1999) and the stripped gas forms sufficient column
density to the central source, it would also work as an X-ray
reflector, albeit ram-pressure stripping is not expected to be
effcient near the centre of a galaxy.

Much of the hard X-ray tail in the southern cone spectrum could be
accounted for by the faint wing of the PSF for the nuclear source,
although a faint extension is present at the inner part of the western
edge of the cone, on inspecting the 4--7 keV image. 
The hard X-ray spectrum lacks a strong Fe K emission line (Fig. 10).
The 90 per cent upper limit on the EW of a narrow line at 6.4
keV is still consistent with the spectral shape
of the nuclear source (Section 3.1). A higher energy line, e.g., at
6.7 keV, appears to be more favoured by the data (in this case, the EW
could be around 800 eV), although the significance of the line is
sensitive to the assumed underlying continuum. This line may arise
from the high ionization gas (log $\xi\sim 3$) in agreement with the
soft X-ray emission (Section 4.1).

\subsection{Extended Fe K$\alpha $ emission}


\begin{figure}
\centerline{\includegraphics[width=0.45\textwidth,angle=0,keepaspectratio='true']{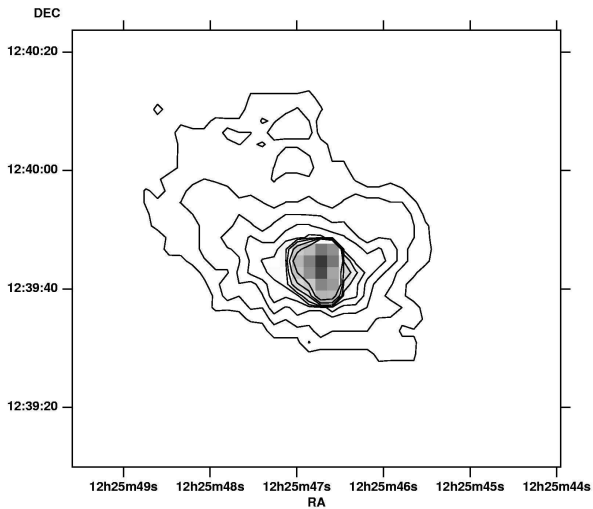}}
\caption{
  The Fe K$\alpha $ band (6.2--6.5 keV) image of NGC4388. The narrow band image
  with 2-arcsec binning has been smoothed and the contours of eight
  logarithmic intervals in the range of 0.05--3 per cent of the peak
  brightness (at the nucleus position) are superposed to show low
  surface brightness morphology. }
\end{figure}

The Fe K$\alpha $ emitting region is spatially resolved in the Chandra
image.  A narrow-band (6.2--6.5 keV) image centred on cold iron
K$\alpha$ emission is shown in Fig. 11. While much of the line
emission arises from the nuclear region, the image of the low
surface-brightness emission shows ridges extending to the north-east
and north-west. A faint extension appears to be present to the north
up to 30 arcsec (2.5 kpc). The reality of this extended line emission
is supported by the hard X-ray emission with an asymmetric extension
examined in the previous subsection (Fig. 9) and the spectrum of the
low surface brightness region (LSB in Fig. 10) which shows a strong Fe
K$\alpha $ line with EW much larger than that of the nuclear spectrum.
The energy of the line centroid ($\simeq 6.4$ keV) of the LSB spectrum
implies that the line emission originates in low ionization gas. The
extension of the line emission is in rough agreement with the
photoionized gas traced by soft X-ray emission, although any detailed
correlation between the two emission components is unclear due to the
low signal to noise ratio of the Fe K$\alpha $ emission data.

The 6.4 keV Fe K$\alpha $ line is a reliable indicator of cold gas,
and could trace the history of nuclear activity by mapping the
iron line emission (Fabian 1977). The extension of the line emission
implies that the central source was turned on at least ten thousands
years ago. More details could be obtained from a deeper observation.

\section{discussion}

The soft X-ray image shows clear bi-conical extended emission,
bearing a strong resemblance to the optical [OIII] ionization cone.
The faintness of the northern cone is probably due to the effect of
shadowing by the stellar disk on the near side (see Section 4.2). The
spatial correlation between the soft X-ray and optical line emission,
which was not found in the much lower signal-to-noise image taken by
the ROSAT HRI (Matt et al 1994), suggests a common mechanism working
for the creation of the extended nebula. As discussed in Section 4.1,
the extended X-ray emission in NGC4388 is unlikely to be thermal
(collisionally ionized) emission from a shock-heated medium, as
previously thought (however, Netzer, Turner \& George 1998 explained
the ASCA spectrum with a photoionization model). Convincing evidence
for photoionized gas that is responsible for the soft X-ray emission
in other Seyfert 2 galaxies has been obtained from high resolution
spectroscopy with the XMM-Newton RGS and the Chandra LETG
(Kinkhabwala et al 2002; Brinkman et al 2002; Ogle et al 2000; Sako et
al 2000; see also Young, Wilson \& Shopbell 2001). This also seems to
be the case for the X-ray nebula around NGC4388. 

The strong OVII emission, together with the other low energy X-ray
emission features, found in the spectrum of the southern cone suggests
that photoionized gas with a low ionization parameter (log $\xi\sim
0.4$) dominates the bright soft X-ray emission. With the inferred low
ionization, the same gas could explain the optical [OIII] luminosity,
since photoionization codes predict the luminosity ratio of
[OIII]$\lambda 5007$ to OVII (0.55 keV) to be $\sim 40$ for the
ionizing spectrum we assumed (this value is from XSTAR). Given the
rough power-law approximation to the shape of the ionizing spectrum
for the photoionization calculation and various uncertainties,
including the aperture difference and varying absorption/extinction
across the region, which complicate any direct comparison, the
observed line flux ratio ($1.2\times 10^{-12}$\ergpspsqcm\ for [OIII]
measured with a $12^{\prime\prime}$ aperture by Falcke et al (1998);
$1.8\times 10^{-14}$\ergpspsqcm\ for OVII measured only from the
southern cone) is in a rough agreement with the prediction. Recall
that the $\xi$ quoted above is for the inner edge of a photoionized gas
shell in the calculation of XSTAR (see Section 4.1), and the
ionization parameter varies across the gas: OVII peaks near the inner
edge while OIII peaks at outer radii where $\xi $ is lower but the
radial emissivity of the two have a significant overlap.

In the southern cone, a comparison between the spectra of the inner
and outer parts suggests the ionization parameter is approximately
constant with radius (Section 4.1). This implies that the radial
density profile follows $n\propto R^{-2}$. Since the temperature
(a few $10^4$ K here) should hold for gas with the same ionization
parameter, the pressure profile is also proportional to the inverse
square of the radius, which is consistent with gas expanding at constant
velocity, ejected matter from the active nucleus, for instance.

If the X-ray nebula around NGC4388 is indeed photoionized gas, as we
demonstrated, its size of over 10 kpc is larger than other
photoionized nebulae known around nearby Seyfert galaxies. High
resolution X-ray spectroscopy is desirable to confirm the
photoionization nature. The large extent of the X-ray emission would
make a Chandra grating observation difficlt. An observation with the
XMM-Newton RGS may still be possible, given the broader PSF of the
XMM-Newton telescopes. Note, however, such a grating observation will
lose information on the spatial variation of gas properties, as we see in
the Chandra data.

Large scale X-ray nebulae with the size up to a few tens kpc, which
are often associated with H$\alpha $ emission (e.g., Heckman et al
1996; Lira et al 2002), are usually attributed to thermal emission
heated by starburst-driven winds (e.g., Lehnert, Heckman \& Weaver
1999; Gallagher et al 2002; Done et al 2003). However, at least in
the presence of an active nucleus, the possibility of photoionized gas
as a source of an extended X-ray nebula should be considered,
especially when there is a correlation with high ionization optical
emission gas such as [OIII]$\lambda 5007$, as seen in NGC4388.

While the radial extent of the southern cone is limited, the question
arises why the extension of the gas to the north is so much larger in
both optical line and X-rays. Various possibilities for the origin of
this gas have been proposed: ram-pressure stripping of the
interstellar medium (Petitjean \& Durret 1993; Veilleux et al 1999),
debris of a small accreted galaxy (Yoshida et al 2001), and a
starburst-driven superwind (e.g., Corbin et al 1988). The last
possibility is not favoured because the X-ray spectrum is not thermal
(see also Veilleux et al 1999; Vollmer \& Huchtmeier 2003). The
extended cold iron K emission rather favours a photoionization model.
The narrow filaments of a few tens of kpc length, imaged by SUBARU,
are certainly suggestive of the merger debris hypothesis. They are
reminiscent of the tidal tails often observed in a major merger
system, e.g., in HI.  Of course, the unusual environment particular to
NGC4388, residing in the cluster core and moving at high velocity in
it, may play an important role in causing the long one-sided matter
distribution.

The radial surface-brightness profile of the soft X-ray emission
suggests that the density profile to the north side of the galaxy is
significantly flatter than that to the south.  If the ionized gas in
the northern extension is excited through photoionization by the
active nucleus, then its luminosity is controlled by the amount of
matter available for photoionization. If the ionization parameters
obtained from the photoionization models of the various regions are
used as a guide, the very low surface brightness region at large radii
(4--9 kpc or NE VLSB; Section 4.4) appears to have a lower ionization parameter
than the inner north plume (within 1 kpc; Section 4.2). The decrease
in the ionization parameter indicates that the density profile is
flatter than $n\propto R^{-2}$. Under a matter-bounded condition, the
emissivity depends on the photoionization rate which is proportional
to density. The observed power-law form of the soft X-ray
surface-brightness profile ($\propto R^{(-1.4\pm 0.1)}$, Section 4.4)
implies that the density profile to the north of the galaxy may have
a similar form, if projection effects are not important, i.e., it has a
shallow depth, for example, of a sheet-like structure, as expected for
a narrow streak of stripped gas, or the debris/tidal tail of a galaxy
merger.

The apparent high ionization condition found at intermediate radii
(LSB; Section 4.3) is not explained by the above scenario. This could
result from a different ionization condition, i.e., a different
distribution function of ionization parameters, in the north-east
``three-shell'' region. Although the X-ray spectrum of that region
alone is too noisy to be conclusive, the X-ray colour map (Fig. 3b)
indicates that the shell region shows a harder spectrum than the
surrounding area. The aligned shell-like structure resembles that in
the optical ionization bi-cone of the Seyfert 2 galaxy NGC5252 (e.g.,
Tadhunter \& Tsvetanov 1989; Acosta-Pulido et al 1996). It is also
reminiscent of the X-ray bow structure found in the X-ray nebula
around the Vela Pulsar (Pavlov et al 2001). These structures may be
due to a density discontinuity created by shock (the shells in NGC5252
may be pre-existing HI rings).  The harder spectrum of the shell
region could be related to the shock-heated media, although increased
absorption is an alternative explanation.

The radio images show that radio-emitting plasma extends to the north
at large radii whereas the axes of the ionization bi-cone and bright
radio emission lie along the NE-SW direction. Veilleux et al (1999)
attributed this misalignment to buoyancy or refractive bending (note
that the NE extraplanar ionized gas is moving towards us and the axis
of the ionization cone is not perpendicular to the stellar disk). As
we noted in Section 4.2, the major axis of the 1--3 keV (green colour
in Fig. 1) or 1--2 keV (Fig. 5) emission, which is likely to trace the
path of the nuclear radiation at inner radii (within 1 kpc), lies in
the NE-SW direction. Its southern extension also match the
well-defined edge of the western side of the southern soft X-ray/OIII
cone. The fact that the eastern edge of the southern cone is more
blurred suggests that some buoyancy effect might bring the ionised gas
southwards, while the true axis of the nuclear radiation which escapes
from the nuclear obscuring torus is indicated by the 1--3 keV
elongation. In this case, although the apparent opening angle of the
southern cone is about $55^{\circ}$, the nuclear ionizing radiation
could be more collimated.  On the north side, both the 1--2 keV X-ray
emitting gas and radio plasma point towards the NE at small radii and bend
towards the north at $\sim 3$ arcsec from the nucleus.

NGC4388 is often believed to be interacting strongly with the Virgo
ICM at supersonic speeds. A Mach cone formed on the far-side of the
galaxy with an opening angle of $\sim 80^{\circ}$ has been proposed by
Veilleux et al (1999). Some temperature rise in the bow-shocked region
around the interaction point is expected above the surrounding ICM,
which should have a temperature of $kT\sim 2$ keV.  Unfortunately, we
were unable to find such a bow-shocked region, mainly due to the
presence of the extended ionized gas. The lack of evidence for strong
interaction with the ICM leaves room for the hypothesis that NGC4388
could be more distant than the Virgo cluster and lies closer to at its
Hubble distance for $cz = 2540 $ \kmps.

Finally, with regard to the unification scheme of the two types of
Seyfert galaxies, if the nucleus of NGC4388 were viewed from a
direction near the axis of the ionization cone, the central X-ray
source would be seen through the low ionization gas (and possibly very
high ionization gas). The column density of the low ionization gas
could be of the order of $10^{22}$\psqcm, depending on the filling
factor.  Although the detailed X-ray grating spectra of bright Seyfert
1 galaxies, such as NGC3783 (Kaspi et al 2002), do not show clear
evidence for such low ionization gas, it might be seen in other
mildly-absorbed Seyfert nuclei.

\section*{Acknowledgements} 

XSTAR is maintained by Tim Kallman and his collaborators at Goddard
Space Flight Center. ACF and KI thank Royal Society and PPARC,
respectively, for support. This research was supported in part by NASA 
through grants NAG 81027 and NAG 81755 to the University of Maryland.

\newpage
\section*{appendix}

Detector-efficiency corrected versions of spectral data shown in Fig.
2, Fig. 4 and Fig. 10 are presented below. The count rate spetra have
been divided by the effective area. The correction is
independent of the fitted model but some uncertainty may have been
introduced due to the detector resolution and data binning. The
time-dependent efficiency degradation in the low energy range specific
to the ACIS-S3 detector, which was taken into account by {\tt acisabs}
in the spectral analysis, has also been included.

\newpage

\begin{figure}
\centerline{\includegraphics[width=0.36\textwidth,angle=270,keepaspectratio='true']{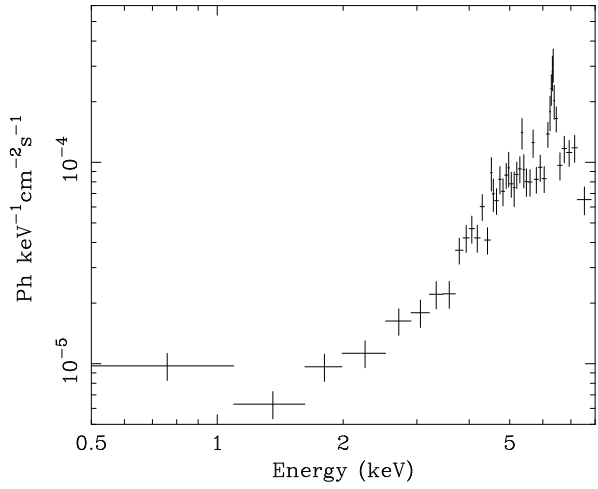}}
\caption{
  The same figures as Fig. 2 but corrected for the detector
  efficiency.}
\end{figure}

\begin{figure}
\centerline{\includegraphics[width=0.6\textwidth,angle=270,keepaspectratio='true']{apfig2.ps}}
\caption{
  The same figures as Fig. 4 but corrected for the detector
  efficiency.}
\end{figure}

\begin{figure}
\centerline{\includegraphics[width=0.55\textwidth,angle=270,keepaspectratio='true']{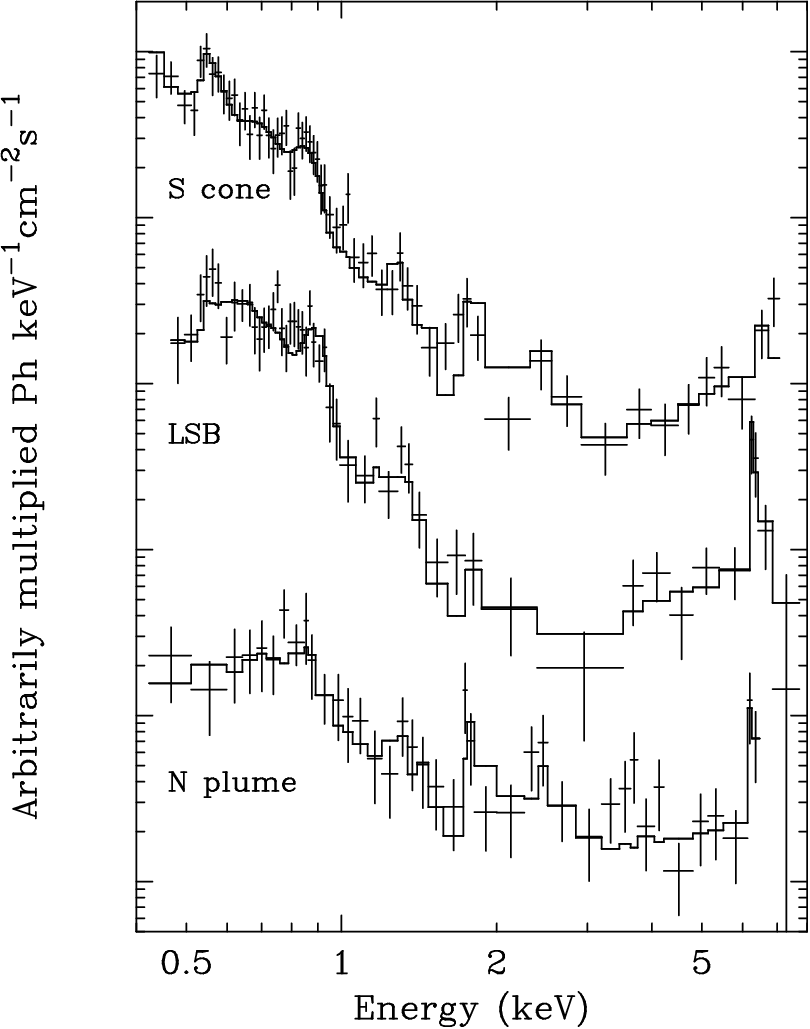}}
\caption{
  The same figures as Fig. 10 but corrected for the detector
  efficiency.}
\end{figure}

\end{document}